\documentclass[conference]{IEEEtran}

\usepackage{cite}
\usepackage{amsmath,amssymb,amsfonts}
\usepackage{graphicx}
\usepackage{textcomp}
\usepackage{booktabs}
\usepackage{algorithm}
\usepackage{algpseudocode}
\usepackage[hidelinks]{hyperref}
\usepackage{orcidlink}

\graphicspath{{figures/}}
\DeclareGraphicsExtensions{.pdf,.png,.jpg,.jpeg}
\usepackage{eso-pic}

\newcommand{\IEEEcopyrightnotice}{
\color{gray}
This work has been submitted to the IEEE for possible publication. Copyright may be transferred without notice, after which this version may no longer be accessible.
\smallskip\par
This manuscript is a preprint and has not yet undergone peer review.
}

\begin{document}

\title{Dual-Parabola Interpolator: A Low-Bias Peak Estimator for Distributed Radar Synchronization}

\author{\IEEEauthorblockN{Matteo Pardi~\orcidlink{0009-0007-0615-6291}\textsuperscript{\dag*},
Giulio Meucci~\orcidlink{0000-0002-5472-2186}\textsuperscript{*},
Francesco Mancuso~\orcidlink{0000-0003-4174-7816}\textsuperscript{*},\\
Anna Lisa Saverino~\orcidlink{0000-0002-9002-3649}\textsuperscript{*},
Amerigo Capria~\orcidlink{0000-0003-4855-5323}\textsuperscript{*}}\\[-1em]
\IEEEauthorblockA{\textsuperscript{\dag}Dept.\ of Information Engineering,
University of Pisa, Pisa, Italy\\
\textsuperscript{*}Radar and Surveillance Systems (RaSS) National Laboratory,
CNIT, Pisa, Italy\\[1em]
\href{mailto:matteo.pardi@phd.unipi.it}{\texttt{matteo.pardi@phd.unipi.it}}}}

\AddToShipoutPictureFG*{
  \AtPageLowerLeft{
    \put(\LenToUnit{(\paperwidth-\textwidth)/2 - 0.05in},\LenToUnit{0.30in}){
      \parbox[b]{\textwidth}{
        \footnotesize\IEEEcopyrightnotice
      }
    }
  }
}

\maketitle

\begin{abstract}
Coherent distributed radar processing requires accurate wireless
synchronization. Two-way time transfer (TWTT) obtains reception timestamps
from sampled matched-filter peaks, but interpolation bias limits accuracy even
when noise is reduced by higher signal-to-noise ratio (SNR) or averaging. Three-point parabolic
interpolation is attractive for its simple, explicit formula; reducing its
bias generally requires more elaborate fitting, such as higher-order
polynomials, or a model tailored to a particular peak shape. A practical
alternative should retain this simplicity and interpretability while
approaching the low bias of more flexible fits. We propose the Dual-Parabola
Interpolator (DPI), which combines two parabolic estimates from five amplitude
samples near the matched-filter peak. An adaptive weight cancels the leading
interpolation error without requiring a waveform-specific peak model or a
correction table. Simulations over flat and standardized multipath channels
show bias comparable to a five-point quartic fit, with essentially the same
standard deviation as the other interpolators and an execution cost close to
a three-point parabolic fit. This combination is useful for TWTT in urban or
other multipath settings, where channel distortion prevents reliance on a
nominal peak shape.
\end{abstract}

\begin{IEEEkeywords}
Multistatic radar, synchronization, two-way time transfer,
Dual-Parabola Interpolator, delay estimation, bias reduction
\end{IEEEkeywords}

\IEEEpeerreviewmaketitle

\section{Introduction}
Coherent operation of distributed radar systems requires a shared time base
among nodes. Accurate synchronization allows bistatic and multistatic
observations to be combined for joint detection, localization, and imaging.
When wired references are impractical, as in mobile or spatially distributed
deployments, this time base must be established over wireless links.

\begin{figure}[!b]
\centering
\includegraphics[width=0.98\columnwidth]{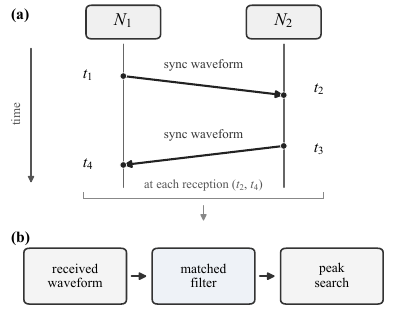}
\caption{Timestamp extraction in two-way time transfer.
(a)~Nodes $N_1$ and $N_2$ exchange synchronization waveforms, transmitted at
$t_1,t_3$ and received at $t_2,t_4$.
(b)~Each reception timestamp is estimated by locating the peak of the matched-filter output
with subsample precision.}
\label{fig:multistatic}
\end{figure}

Two-way time transfer (TWTT) provides a practical means of recovering clock
offsets and propagation delays from exchanged synchronization
waveforms~\cite{prager2020,mghabghab2022,merlo2023}. At each reception, a
matched filter produces a correlation peak whose position supplies the
timestamp (Fig.~\ref{fig:multistatic}). Since the required timing precision
is finer than the sampling interval, the discrete maximum must be refined by
subsample interpolation. Noise causes timestamp fluctuations, whereas the
mismatch between an interpolation model and the actual peak produces a
systematic error that depends on the peak's position relative to the sampling
instants. Higher signal-to-noise ratio (SNR) and averaging reduce the fluctuations, but a repeated
fractional offset retains this interpolation bias and sets an accuracy floor.

Three-point parabolic interpolation remains an appealing choice because its
closed-form vertex is inexpensive to compute and easy to interpret. Its
fractional-delay bias is well established for sampled
correlators~\cite{boucher1981,moddemeijer1991,jacovitti1993,jameson2006}.
Fitting a parabola to the log-magnitude is similarly simple and is used in
TWTT~\cite{prager2020} and Gaussian peak localization~\cite{guo2011}, but it
also leaves a residual when the transformed peak is not quadratic. More
flexible local models, such as splines or higher-order polynomials, can
reduce this mismatch, at the cost of additional fitting and higher complexity. Bias can also be controlled through reconstructive
interpolation~\cite{cespedes1995} or by adjusting the spacing of auxiliary
samples in spectral peak estimation~\cite{belega2021,togni2025}.

An alternative is to exploit a known peak shape. Sinc nonlinear least squares
(NL-LS) has been used for radar synchronization~\cite{prager2020,kenney2024},
while matched-filter-bank least squares uses modeled correlation responses
for ranging~\cite{mghabghab2022}. Analytic or tabulated corrections can also
compensate parabolic bias~\cite{cespedes1995,lai1999}; quadratic fits followed
by lookup-table correction have enabled picosecond TWTT and decentralized
synchronization~\cite{merlo2023,shandi2025}. Such methods are effective when
their model describes the received peak. In urban or other multipath
settings, however, the channel can deform that peak, making a correction
tailored to the nominal waveform less reliable.

The practical challenge is therefore to lower interpolation bias while
retaining the simplicity of a parabolic fit and avoiding dependence on a
specific received peak shape. Simple formulas are valuable both for
lightweight processing at every exchange and for understanding how local
sample distortions affect the estimate. This motivates the estimator proposed
here. The main contributions of this paper are:
\begin{itemize}
\item We introduce the Dual-Parabola Interpolator (DPI), a closed-form combination of two parabolic estimates from five samples,
with an adaptive weight inspired by Richardson
extrapolation~\cite{richardson1927}. A local peak expansion explains the
cancellation of the leading error and gives the residual bias.
\item We perform a simulation comparison over flat, EPA, and ETU channels showing that
DPI combines the low bias of more elaborate five-sample fits with a cost close
to elementary three-point interpolation, while maintaining similar noise
sensitivity.
\end{itemize}
Section~\ref{sec:dpi} derives the DPI estimator, Section~\ref{sec:results}
presents the simulations, and Section~\ref{sec:conclusion} concludes the paper.

\section{Dual-Parabola Interpolator}
\label{sec:dpi}
\begin{figure}[!t]
\centering
\includegraphics[width=0.98\columnwidth]{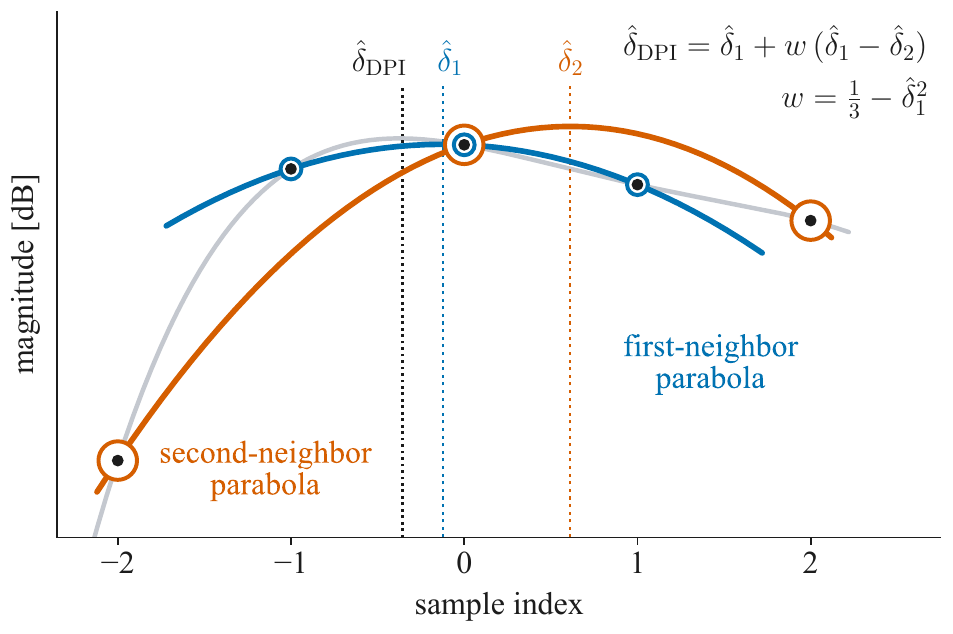}
\caption{DPI combines the vertices of two parabolas fitted to the five values
$y_j$ of a peak (gray curve). The blue parabola uses $y_{-1},y_0,y_{+1}$;
the orange parabola uses $y_{-2},y_0,y_{+2}$. The adaptive combination of their
vertices, $\hat\delta_1$ and $\hat\delta_2$, gives $\hat\delta_{\mathrm{DPI}}$.}
\label{fig:dpi_process}
\end{figure}

The aim of DPI is to reduce the interpolation bias that remains at high SNR.
We therefore derive it for a noise-free peak; its sensitivity to noise is then
evaluated by simulation. Let $z(t)$ be the complex matched-filter output,
$\tau$ its continuous peak location (unknown), and $t_0$ the sampled maximum. Denote by
$f_s$ the acquisition sampling rate and by $h$ the effective interval between
the samples used for interpolation; without sample refinement, $h=1/f_s$,
whereas if classical frequency-domain zero padding is used, $h$ can be reduced below $1/f_s$
(Section~\ref{sec:setup}).
The unknown to be estimated is the normalized offset $\delta=(\tau-t_0)/h$,
with $\delta\in[-0.5, 0.5]$ the central interval of interest around a correctly acquired
sampled peak. The five neighboring samples, expressed in decibels, are
\begin{equation}
  y_j=10\log_{10}|z(t_0+jh)|^2,
  \qquad j\in\{-2,-1,0,1,2\}.
  \label{eq:five_samples}
\end{equation}
These samples must be close enough to describe the same smooth peak lobe,
requiring a sufficiently small sampling interval $h$.
Two parabolas are fitted about the same
center: one uses the nearest neighbors and the other the next-nearest
neighbors (Fig.~\ref{fig:dpi_process}). Their vertices, both expressed in
units of $h$, are
\begin{equation}
  \hat\delta_1=\frac{y_{+1}-y_{-1}}{2(2y_0-y_{-1}-y_{+1})},
  \quad
  \hat\delta_2=\frac{y_{+2}-y_{-2}}{2y_0-y_{-2}-y_{+2}}.
  \label{eq:q1q2}
\end{equation}

If the five values follow an exact parabola, both estimates equal $\delta$.
For a general smooth peak, their errors differ because the fits use spacings
$h$ and $2h$. DPI exploits this difference to cancel the leading error,
following the principle of Richardson extrapolation~\cite{richardson1927}.
To determine the appropriate combination, define the continuous local profile
$g(u)=10\log_{10}|z(\tau+u)|^2$. Assume that its local departure from a
parabola is well described by a fourth-order polynomial, allowing for both
symmetric and asymmetric deformations:
\begin{equation}
  g(u)=g_0-Au^2+A\alpha u^3+A\beta u^4+O(u^5),
  \label{eq:local_model}
\end{equation}
where $A>0$, $\alpha$ describes the leading asymmetry, and $\beta$ the
quartic deformation. The level $g_0$ and scale $A$ cancel in the vertex
formulas, so neither affects the design of the combination. Substituting
$y_j=g((j-\delta)h)$ into~\eqref{eq:q1q2} and taking a Taylor expansion for
small $h$, with $\delta$ fixed, gives
\begin{equation}
\begin{aligned}
  \hat\delta_1&=\delta-\tfrac{\alpha h}{2}(3\delta^2-1)+O(h^2),\\
  \hat\delta_2&=\delta-\tfrac{\alpha h}{2}(3\delta^2-4)+O(h^2).
\end{aligned}
\label{eq:q_expand}
\end{equation}
Thus, asymmetry introduces an $O(h)$ error in each normalized offset. This
term is absent for a symmetric peak, but channel distortion can introduce
asymmetry, so symmetry cannot generally be assumed. The known dependence on
sample spacing makes cancellation possible without
estimating the unknown coefficient $\alpha$.

In particular, $\hat\delta_1-\hat\delta_2=-3\alpha h/2+O(h^2)$, so adding
$(1/3-\delta^2)(\hat\delta_1-\hat\delta_2)$ to $\hat\delta_1$ cancels its
$O(h)$ term. Replacing the unknown $\delta$ by the available estimate
$\hat\delta_1$ yields the implementable rule
\begin{equation}
  \hat\delta_{\mathrm{DPI}}=\hat\delta_1+w(\hat\delta_1-\hat\delta_2),
  \quad w=\tfrac{1}{3}-\hat\delta_1^2,
  \label{eq:weights}
\end{equation}
and the timestamp estimate
\begin{equation}
  \hat\tau_{\mathrm{DPI}}=t_0+h\hat\delta_{\mathrm{DPI}}.
  \label{eq:DPI}
\end{equation}
Since $\hat\delta_1=\delta+O(h)$, this substitution changes the combination
only at $O(h^2)$. Thus $\hat\delta_{\mathrm{DPI}}-\delta=O(h^2)$, and the
timing error drops from $O(h^2)$ to $O(h^3)$ for asymmetric peaks.
Alg.~\ref{alg:dpi} summarizes the two parabolic interpolations and their
weighted combination.

\begin{algorithm}[!t]
\caption{Dual-Parabola Interpolator (DPI)}
\label{alg:dpi}
\begin{algorithmic}[1]
\Require Matched-filter output $z(t)$, sampling interval $h$,
and sampled peak location $t_0$
\Ensure $\hat\tau_{\mathrm{DPI}}$
\State $y_j\gets10\log_{10}|z(t_0+jh)|^2,\quad j\in\{-2,-1,0,1,2\}$
\State $\displaystyle\hat\delta_1\gets\frac{y_{+1}-y_{-1}}{2(2y_0-y_{-1}-y_{+1})}$
\State $\displaystyle\hat\delta_2\gets\frac{y_{+2}-y_{-2}}{2y_0-y_{-2}-y_{+2}}$
\State $w\gets1/3-\hat\delta_1^2$
\State $\hat\delta_{\mathrm{DPI}}\gets\hat\delta_1+w(\hat\delta_1-\hat\delta_2)$
\State $\hat\tau_{\mathrm{DPI}}\gets t_0+h\hat\delta_{\mathrm{DPI}}$
\end{algorithmic}
\end{algorithm}

A closed-form expression for the leading residual bias can be obtained by taking the Taylor
expansion of the two offsets~\eqref{eq:q1q2} through $h^2$, with remainder $O(h^3)$, using the
same local model~\eqref{eq:local_model}.
Substituting these expansions into~\eqref{eq:weights}, including the adaptive
weight, and multiplying the normalized error by $h$ as in~\eqref{eq:DPI} gives
\begin{equation}
  \hat\tau_{\mathrm{DPI}}-\tau
  =h^3\left[\tfrac{3}{2}\alpha^2\delta
    +\left(\beta-\tfrac{9}{2}\alpha^2\right)\delta^3\right]+O(h^4).
  \label{eq:bias_shot}
\end{equation}
Both $\alpha$ and $\beta$ affect the residual. It is worth noting that the leading
term is odd in $\delta$. Therefore, a symmetric distribution of fractional
offsets across transmissions makes the mean leading residual zero.
This suggests future TWTT experiments with known transmit-time
dither, varied between messages and subtracted from the recovered timestamps,
to reduce the mean leading residual.

\section{Results}
\label{sec:results}
\subsection{Simulation Setup}
\label{sec:setup}
The simulations assess how interpolation accuracy changes when the received
correlation peak departs from its nominal shape. We start with a flat channel,
where a sinc-based fit is well matched to the peak, and introduce multipath
to test sensitivity to channel-induced deformation. All estimators receive
samples of the same matched-filter output, so the comparison isolates the
interpolation stage of the TWTT timestamp.

\emph{Waveform and sampling.} The synchronization waveform is an LFM chirp of bandwidth $B=20$~MHz and
duration $5~\mu$s, acquired in complex baseband at $f_s=B$, giving $N=100$
samples. Its matched-filter peak is approximately
sinc-shaped in the flat channel. To obtain a finer temporal sampling grid,
the matched filter is applied in the frequency domain and its output spectrum
is zero-padded before transformation back to time. With zero-padding factor
$\mathrm{ZP}$, the effective sampling interval becomes
$h=1/(\mathrm{ZP}f_s)=1/(\mathrm{ZP}B)$. We vary
$\mathrm{ZP}\in\{4,8,16,32\}$; the minimum factor of four keeps the five-sample
neighborhood sufficiently close to the peak for the local approximation in
this setup.

\begin{figure}[!t]
\centering
\includegraphics[width=\columnwidth]{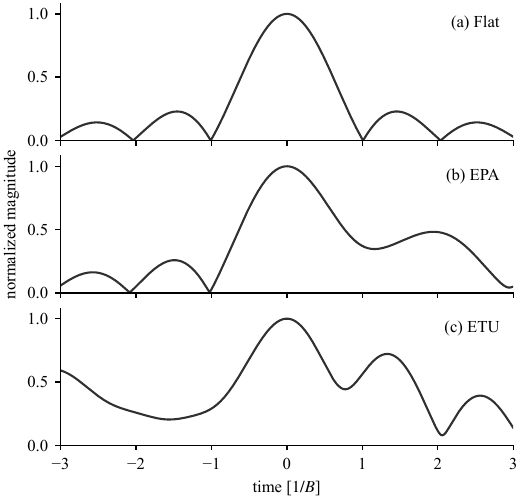}
\caption{Illustrative matched-filter output, normalized and
centered at their continuous maxima: (a)~flat channel, (b)~EPA realization~0 (one of the 100 realizations),
and (c)~ETU realization~77 (one of the 100 realizations). The selected examples show increasing deformation
of the nominal sinc-like peak.}
\label{fig:channel_peaks}
\end{figure}

\emph{Channels.} To assess sensitivity to channel-induced deformation, we also use the
standardized Extended Pedestrian~A (EPA) and Extended Typical Urban (ETU)
multipath profiles~\cite{3gpp36104}. EPA primarily introduces unresolved echoes,
whereas ETU includes more widely separated paths. Fig.~\ref{fig:channel_peaks}
illustrates how these channels can deform the main peak and introduce
nearby maxima. We use one flat-channel case and 100 independent Rayleigh
realizations per multipath profile, each held static during its trials.
For TWTT, a common channel-induced peak shift cancels from the clock-offset
estimate when the link is reciprocal and static over the exchange.
We therefore take as ground truth the continuous maximum of the
noise-free matched-filter output after the channel, including its deformation.
Interpolation errors remain because the peak generally occupies different
fractional positions relative to the two nodes' sampling instants.

\emph{Estimators.} We compare seven local interpolators. The elementary three-point methods
are a parabola on $|z|$~\cite{boucher1981,cespedes1995}, a
log-parabola~\cite{prager2020}, and a cosine fit~\cite{cespedes1995}.
Sinc NL-LS fits three magnitude samples with a free main-lobe width using up
to five Gauss--Newton iterations~\cite{prager2020}. The five-point alternatives
are a cubic spline on the magnitude~\cite{mghabghab2022}, a fourth-order
polynomial (quartic) fitted to the five log-samples, and DPI. (For the quartic, after locating
the local maxima, we select the one nearest to the sampled peak
within one effective sampling interval.) These methods span simple
vertex formulas, shape-specific iterative fitting, and more flexible local
interpolation.

\emph{Delay and noise.} For each channel realization, we delay the waveform over $[-0.5/B,0.5/B]$
in steps of $0.05/B$ to uniformly sample fractional peak positions $\delta$ within a
sampling interval $[-0.5, 0.5]$. White complex Gaussian input noise
produces correlated noise at the matched-filter output. The matched-filter
output SNR ranges from 5 to
40~dB in 0.5~dB steps, with 200 noise realizations per channel realization and
delay. Acquisition is restricted to within $2/B$ of the true peak; boundary
hits and errors larger than $0.5/B$, corresponding to selection of another
maximum, are excluded from the interpolation statistics.

\emph{Metrics.} To evaluate interpolation performance, first define the estimation error of an
individual trial:
\begin{equation}
  e_{mcrkn}(\mathrm{ZP},\mathrm{SNR})
  =\hat\tau_{mcrkn}(\mathrm{ZP},\mathrm{SNR})-\tau_{crk},
  \label{eq:error_metric}
\end{equation}
where $m$ denotes the interpolator, $c\in\{\mathrm{flat},\mathrm{EPA},\mathrm{ETU}\}$
the channel profile, $r$ its realization, $k$ the imposed physical delay
(associated with true offset $\delta$), and $n$ the noise realization.
The ground truth $\tau_{crk}$ is the continuous noise-free peak after the
channel and imposed delay.

Our main interest is the residual bias at high SNR. Since in TWTT the two nodes
generally have different fractional offsets, we assess the worst case over
the tested physical delays rather than averaging away this dependence.
We then average the absolute noise-free error $e^{(0)}$ over channel
realizations:
\begin{equation}
  b_{mc}(\mathrm{ZP})=
  \left\langle\max_k|e^{(0)}_{mcrk}(\mathrm{ZP})|\right\rangle_r,
  \label{eq:bias_metric}
\end{equation}
where $\langle\cdot\rangle_r$ denotes the arithmetic mean over index $r$.
Noise sensitivity is measured by the standard deviation across noise
trials at each fixed condition, followed by the same maximum and average:
\begin{equation}
  \sigma_{mc}(\mathrm{ZP},\mathrm{SNR})=
  \left\langle\max_k\operatorname{STD}_n[e_{mcrkn}]\right\rangle_r.
  \label{eq:std_metric}
\end{equation}
The operator $\operatorname{STD}_n$ takes the standard deviation over
noise realizations (index $n$), keeping all other indices and arguments fixed.
The residual bias~\eqref{eq:bias_metric} and noise sensitivity~\eqref{eq:std_metric}
characterize each interpolator--channel pair. For an overall comparison, we
average equally over the three channels:
$b_m=\langle b_{mc}\rangle_c$ and
$\sigma_m=\langle\sigma_{mc}\rangle_c$, at fixed ZP and SNR.

Execution times refer to single-thread, numba-compiled implementations,
using the minimum of 15 runs of $2\cdot10^5$ estimates. The computational costs
of matched filtering and zero padding are common to all compared methods and
are excluded so that the benchmark measures the interpolator alone.

\subsection{Bias}
\begin{figure}[!t]
\centering
\includegraphics[width=\columnwidth]{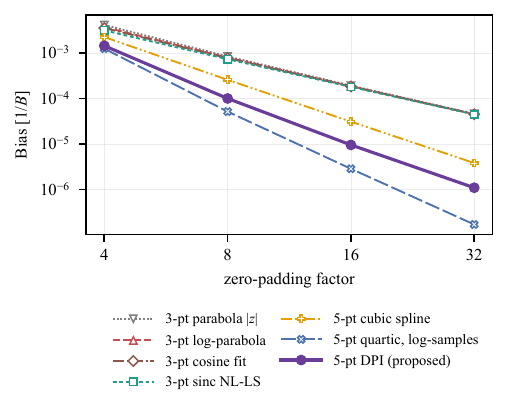}
\caption{Bias $b_m(\mathrm{ZP})$ versus zero-padding factor $\mathrm{ZP}$.
The bias is averaged over the simulated channels.
The quartic fit and DPI achieve the lowest bias.}
\label{fig:bias_vs_zp}
\end{figure}

Fig.~\ref{fig:bias_vs_zp} shows the aggregated bias $b_m(\mathrm{ZP})$. As
expected, a finer temporal sampling grid reduces the error for every method.
The elementary three-point
fits form a group with higher residual bias, and the sinc fit joins this
group once results are averaged across channel profiles. The more flexible
five-point methods separate from this group: the spline reduces the bias,
while DPI and the quartic achieve the lowest values. At $\mathrm{ZP}=8$, DPI reaches
$1.0\cdot10^{-4}/B$ (5~ps), about ten times below the three-point
fits. Its bias is comparable to the quartic's, although the quartic remains
more accurate. The significance of this comparison becomes clear when the
execution cost is considered in Section~\ref{sec:cost}.

\begin{figure}[!t]
\centering
\includegraphics[width=\columnwidth]{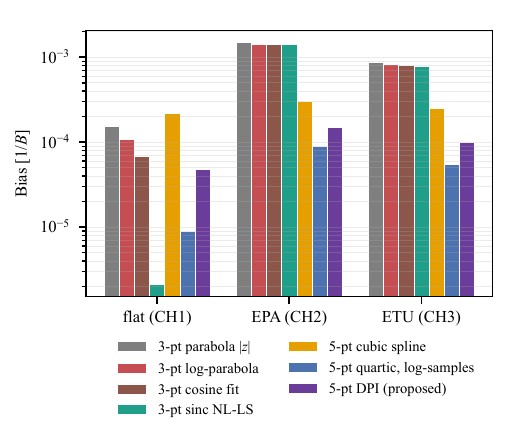}
\caption{Bias $b_{mc}(\mathrm{ZP})$ versus channel profile.
Here, $\mathrm{ZP}=8$ is used as a representative setting.
The sinc NL-LS advantage in the flat channel is lost under multipath
deformation, where the quartic fit and DPI achieve the lowest bias.}
\label{fig:bias_channel}
\end{figure}

Fig.~\ref{fig:bias_channel} separates the channel profiles at $\mathrm{ZP}=8$.
The sinc NL-LS has a particularly small bias on the flat channel, consistent
with its fit to the nominal peak shape. Its advantage disappears when
multipath deforms that shape: on EPA and ETU, its bias is close to that of the
elementary three-point fits. DPI, the spline, and the quartic retain lower
bias on these channels, with the quartic giving the smallest values. This
contrast motivates a local estimator that does not rely on an exact sinc
shape: DPI approaches the accuracy of the more elaborate fits using only
two parabolic vertices and their explicit combination.

\subsection{Standard Deviation}
\begin{figure}[!t]
\centering
\includegraphics[width=\columnwidth]{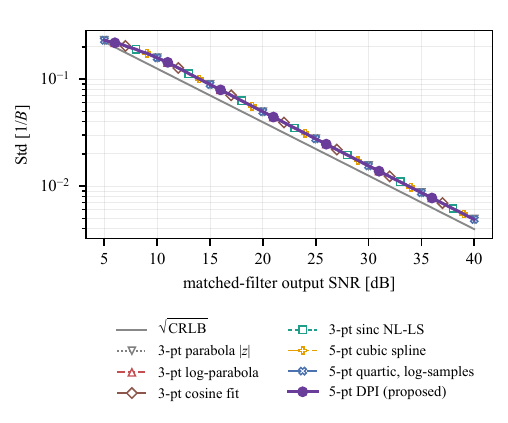}
\caption{Standard deviation $\sigma_m(\mathrm{ZP},\mathrm{SNR})$ of a single
estimate versus matched-filter output SNR. Here, $\mathrm{ZP}=8$ is used as
a representative setting; the reference is $\sqrt{\mathrm{CRLB}}$ for the
nominal waveform. All estimators exhibit similar standard deviations, indicating
that their main differences lie in the bias.}
\label{fig:std}
\end{figure}

Fig.~\ref{fig:std} reports $\sigma_m(\mathrm{ZP}=8,\mathrm{SNR})$ as a
representative case. The curves nearly overlap, lying slightly
above the square root of the Cram\'er--Rao lower bound (CRLB) for the nominal
waveform. This bound provides an ideal reference; the multipath cases use a
template matched to the transmitted waveform rather than to the distorted
received signal. The interpolators therefore have similar noise sensitivity
under the evaluated conditions; their main difference is the residual bias.

\subsection{Computational Cost}
\label{sec:cost}
\begin{figure}[!t]
\centering
\includegraphics[width=\columnwidth]{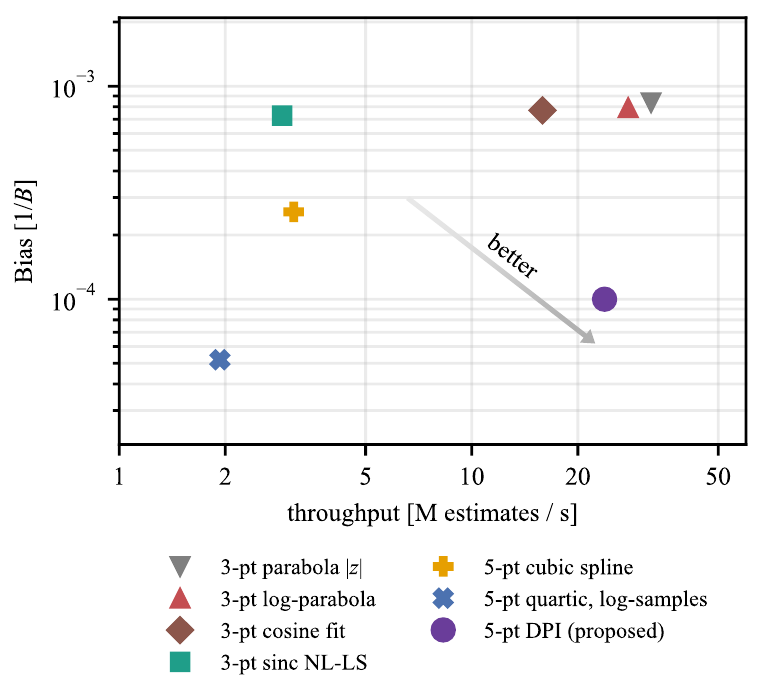}
\caption{Bias $b_m(\mathrm{ZP})$ versus throughput (the inverse of execution time
per estimate). Here, $\mathrm{ZP}=8$ is used as a representative setting.
DPI achieves low bias comparable to more elaborate fits,
such as the quartic, while retaining simplicity and computational cost
comparable to three-point parabolic interpolation.}
\label{fig:throughput}
\end{figure}

Table~\ref{tab:results} and Fig.~\ref{fig:throughput} show how the bias
reduction translates into a practical accuracy--cost trade-off. The elementary
three-point fits are fast but retain a higher bias on distorted peaks.
Sinc NL-LS is more expensive and provides its main benefit when the assumed
shape holds. Spline and quartic fitting reduce bias under distortion, but
require additional calculations to locate a continuous maximum.

DPI occupies a different position: it combines the low-bias behavior of the
five-point fits with execution time close to the elementary methods. In the
measured implementations it takes 42~ns per estimate, compared with 36~ns for
the log-parabola, 320~ns for the spline, and 518~ns for the quartic. Relative
to a log-parabola, DPI adds only two logarithms, a second vertex, and a scalar
combination. The quartic achieves a lower bias, but is about ten times
slower here. These measurements support the intended role of DPI as a simple
interpolator for frequent TWTT exchanges, with low bias and transparent
sample-to-estimate calculations.

\begin{table}[!t]
  \centering
  \caption{Measured Execution Time and Bias}
  \label{tab:results}
  \begin{tabular}{@{}lccc@{}}
  \toprule
  Method & Time & \multicolumn{2}{c}{Bias [$10^{-4}/B$]} \\
  \cmidrule(lr){3-4}
   & [ns] & ZP $=4$ & ZP $=8$ \\
  \midrule
  3-pt parabola $|z|$~\cite{boucher1981,cespedes1995} & \underline{31} & 41.1 & 8.38 \\
  3-pt log-parabola~\cite{prager2020} & 36 & 36.8 & 7.91 \\
  3-pt cosine fit~\cite{cespedes1995} & 63 & 35.1 & 7.70 \\
  3-pt sinc NL-LS~\cite{prager2020} & 345 & 31.3 & 7.28 \\
  5-pt cubic spline~\cite{mghabghab2022} & 320 & 22.3 & 2.57 \\
  5-pt quartic, log-samples & 518 & \underline{12.4} & \underline{0.52} \\
  \textbf{5-pt DPI (proposed)} & \textbf{42} & \textbf{14.4} & \textbf{1.00} \\
  \bottomrule
  \multicolumn{4}{@{}p{0.9\columnwidth}@{}}{\footnotesize Bias: $b_m(\mathrm{ZP})$, averaged over the three channels.
  Bold: proposed method; underlined: best value in each column.
  DPI achieves bias comparable to more elaborate fits (e.g., quartic), with
  simplicity and cost comparable to three-point parabolic interpolation.}
  \end{tabular}
\end{table}

\section{Conclusion}
\label{sec:conclusion}
We proposed the Dual-Parabola Interpolator, a new estimator for subsample
peak timing in distributed radar synchronization. Its explicit combination
of two parabolic estimates reduces
interpolation bias while preserving an elementary calculation. Over the
simulated flat and multipath channels, DPI achieves bias comparable to a
five-point quartic fit, with similar noise-induced dispersion and a cost close
to three-point parabolic interpolation. Its independence from a specific peak
shape supports low-bias operation across the evaluated channel deformations.
The five samples must remain close
to the peak, requiring a zero-padding factor of at least four in the evaluated
setup. Future work will investigate known temporal dithering to reduce the
mean leading residual over TWTT exchanges and experimental validation on
software-defined radar nodes.

\section*{Acknowledgment}
The authors used Claude (Anthropic) and ChatGPT (OpenAI) to assist in the preparation
of this paper: to edit the text of all sections, to develop and check the simulation
code used for Section~\ref{sec:results}, and to produce the plotting scripts for the
figures in Section~\ref{sec:results}. All content was reviewed and verified by the
authors, who take full responsibility for it.
\bibliographystyle{IEEEtran}
\bibliography{references}

\end{document}